\documentstyle[aps,prl,manuscript]{revtex}
\begin{document}

\title{Chirality of internal metallic and semiconducting carbon nanotubes}
\author{R.R. Bacsa$^{2}$, A. Peigney$^{2}$, Ch. Laurent$^{2}$, P. Puech$^{1}$, W.S. Bacsa $^{1}$ \cite{email}}
\address{$^{1}$ Laboratoire de Physique des Solides, UMR CNRS 5477, L'Institut de Recherche sur 
les Systèmes Atomiques et Moléculaires Complexes, Universit\'e Paul Sabatier, 118 route de Narbonne, 
31062 Toulouse C\'edex 4, France }
\address{$^{2}$ Centre Interuniversitaire de Recherche et d'Ingénierie des Matériaux, UMR CNRS 5085,
Universit\'e Paul Sabatier, 118 route de Narbonne, 31062 Toulouse C\'edex 4, France}

\maketitle

\begin{abstract}

We have assigned the chirality of the internal tubes of double walled carbon nanotubes grown by catalytic 
chemical vapor deposition using the high sensitivity of the radial breathing mode (RB) in inelastic light 
scattering experiments. The deduced chirality corresponds to several semiconducting and only two metallic 
internal tubes. The RB modes are systematically shifted to higher energies when compared to theoretical 
values. The difference between experimental and theoretical energies of the RB modes of metallic tubes 
and semiconducting tubes are discussed interms of the reduced interlayer distance between the internal 
and the external tube and electronic resonance effects. We find several pairs of RB modes corresponding 
to different diameters of internal and external tubes.

\end{abstract}



Carbon nanotubes have unique properties due to their low dimensionality \cite{msdress}. Numerous potential 
applications depend critically not only on the control of their chirality, diameter, number of layers and 
tube length but also on the development of high yield and low cost synthesis techniques \cite{rbac}. It has 
been recently shown that the catalytic chemical vapor deposition (CCVD) technique based on transition metal 
oxides produces small diameter carbon nanotubes having $1-4$ walls with high yield and low cost \cite{rbac}. 
The electronic and optical properties of the tubes are linked to their chirality and they can be semiconducting 
as well as metallic \cite{msdress,hamada,wildoer,rinzler,tasaki}. Metallic carbon nanotubes are considered to 
be model systems for one dimensional electronic transport where strong electronic correlations lead to 
collective low energy excitations \cite{egger,kane}. Recent experimental reports suggest the existence of 
superconductivity in bundles of single walled carbon nanotubes \cite{bouchiat}. Double walled carbon nanotubes 
(DWCNT) with an internal metallic tube and an insulating external tube could represent the ultimate one dimensional 
carbon based conductor. For DWCNT the chirality of the external tube can in principle be determined by scanning 
probe techniques or by electron or x-ray diffraction \cite{lieber,he}. However, the determination of 
the chirality of the internal tube remains particularly challenging and has not been determined so far. 
The energy of the radial breathing (RB) mode in carbon nanotubes is inversely proportional to its diameter 
\cite{band} which gives the possibility to find the diameter distribution using inelastic light scattering 
provided the scattering crosssection does not vary strongly with the tube diameter. For tube diameters smaller 
than $1nm$, the sensitivity of the radial breathing mode (RB) to the diameter is particularly high and ranges 
between $220-470cm^{-1}$/nm for a tube diameter between $1.0-0.7nm$ \cite{band}. The chirality of the tubes 
can be explained by the fact that the honeycomb lattice can have different orientations with respect to the 
tube axis. For a given diameter range the number of different orientations possible for a honeycomb lattice 
is restricted to a finite number and can be found by the chiral vector specified by $(n,m)$ \cite{msdress}. 
The chiral vector connects equivalent points in the honeycomb lattice; its length determines the tube 
circumference and its direction gives the orientation with respect to the tube axis. For smaller diameters 
the number of unit cells forming the circumference decreases and hence the number of tubes with different 
chirality within a given diameter range decreases. Combined with the higher sensitivity of the RB mode it 
becomes possible, provided the the observed RB modes are sufficiently narrow, to identify from the position 
of the RB mode the chirality of the tube .



Carbon nanotubes were prepared by the catalytic chemical vapor deposition method (CCVD) \cite{rbac}. Selective 
reduction at $1000^{o}C$ in a methane-hydrogen $(18\% CH_{4})$ atmosphere of a solid solution of a transition 
metal oxide and an insulating oxide($Mg_{0.95}Co_{0.05}O$) led to the formation of small diameter carbon 
nanotubes with $1-4$ walls. After the reaction, the unreacted Co and MgO were dissolved in dilute ($3.7\%$) 
hydrochloric acid and carbon nanotubes were recovered. High resolution electron microscopy images showed 
the presence of individual and bundles of nanotubes of diameters ranging from $0.6-3nm$. Most of them were 
single or double walled and were $>100\mu m$ in length. Chemical analysis showed that the sample contained 
$~95 vol.\%$ of carbon. 

We have selected a sample with mostly single and double walled tubes as verified by high resolution 
electron microscopy \cite{stat} with a significant fraction of internal tubes with diameters smaller than 
$1nm$ and we have examined the inelastic light spectrum of tubes along a single tube bundle. The tubes 
agglomerate to networks of bundles or ropes when deposited on a silica surface from a suspension 
\cite{rbac,ajayan}. We find that the packing of ropes depends on the diameter of the single and double 
walled tubes. The bundles have a variable diameter due to fact that they consist of tubes with different 
diameter and can on some locations consist of only a few tubes or one single tube.
We have dispersed the tubes on a bilayer substrate to enhance the light scattering signal due 
to the enlarged local field \cite{wbac}. This enhancement technique has the advantage in that it is not based 
on electronic resonance which can modify the spectra. We have used a micro-Raman spectrometer (DILOR XY, 
objective x100, spot size: $0.7\mu m$) with a micro-displacement stage to record successive spectra in steps 
of $0.5\mu m$ along a single tube bundle.



Raman scattering from vibrational modes of single wall carbon nanotubes have been reported earlier and 
have provided valuable structural and electronic information \cite{richter,rao,wbac1}. Strong in-plane covalent 
bonds and weak inter-planar interaction result in several in-plane modes in the range $1560-1600cm^{-1}$, 
a disorder induced band in the range $1300-1400cm^{-1}$ and out-of-plane (radial) modes at $<400cm^{-1}$. The 
mode with the highest intensity at low energy is the radial breathing mode with symmetry $A_{1g}$. Modes of 
symmetry $E_{2g}$ at low energy have not been observed so far due to their presumably low intensity \cite{saito}. 
Single wall CNT's grown by other techniques such as the pulsed laser evaporation \cite{thess} or the 
electric arc technique \cite{journet} have a typical diameter of $1.35nm$ and a RB mode at $157cm^{-1}$. 
We present here inelastic light spectra from DWCNT grown by the CCVD method with a significant fraction of 
tubes with small internal diameter tubes and RB modes between $150-370cm^{-1}$. While the electronic 
structure depends on tube diameter and chirality, the light scattering resonance is expected to follow the 
changes in the electronic structure of the tube. We have selected the 488nm line of the argon laser which 
is in resonance with metallic single walled tubes of diameter 0.95nm and semiconducting tubes of diameter 
$1.2nm$ \cite{wang}. All the tube-diameters have been determined by using the inverse relationship between 
diameter and RB energy \cite{band}. 



Figure 1 shows spectra on two location on the sample in the energy range of the RB modes. 
The spectrum at the bottom shows a RB mode at $204cm^{-1}$ which corresponds to a diameter of 1.096nm 
and can be assigned to single walled tubes. Tubes with several different chiralities fall within the 
interval of the peakwidth and an assignment to one chirality cannot be made. The non-homogeneous 
line-shape is an indication that several diameters contribute to the observed peak. The spectrum at the top 
with two peaks at $167cm^{-1}$ and $302cm^{-1}$ is consistent with DWCNTs with external and 
internal diameters of 1.340nm and 0.741nm. The diameter of the internal tube is close to the diameter 
(0.747nm) which corresponds to the chirality (9,1) and (6,5). The two chiralities have by accident the 
same diameter and are both predicted to be semiconducting \cite{msdress}. There are tubes with chirality 
(7,4) and (8,2) wih a similar diameter and they may also contribute to the observed band at 
$302cm^{-1}$. From the internal and the external diameter we find that the deduced interlayer 
distance (0.30nm) is smaller than in graphite (0.336nm). 

We recorded several series of spectra along single bundles and we show here three spectra (figure 2) out of 
50 along one tube bundle which have RB modes in the $280-350cm^{-1}$ range corresponding to tubes with 
particularly small diameter. We have indicated the calculated RB energies for all possible chiralities 
near the horizontal axis. The number of RB mode energies gets considerably smaller above $250cm^{-1}$. 
The spectrum at the bottom shows three main bands at $166cm^{-1}$, $202cm^{-1}$ and 
$300cm^{-1}$ with corresponding diameters of $1.35nm$, $1.1nm$ and $0.741nm$. 
The three peaks at $300cm^{-1}$ are narrower than in figure 1 and can be assigned to tubes of chirality 
(7,4), (9,1), (6,5) and (8,2). The tubes with chirality (7,4) and (8,2) both predicted to be metallic 
\cite{msdress} fall within $1cm^{-1}$ of the predicted values (Table I). It is interesting to note that 
the peak at $302cm^{-1}$ which corresponds to two different chiralities both predicted to be 
semiconducting, has a larger intensity and is shifted to larger energies. We can pair the RB modes at 
$300cm^{-1}$ with RB mode at 166cm$^{-1}$ and we deduce an interlayer spacing of $0.297-0.316nm$. In the 
spectrum in the middle of figure 2 we observe several RB modes between $150cm^{-1}$ and $343cm^{-1}$. We can 
assign the modes at $251cm^{-1}$ and $150cm^{-1}$ to double walled tubes corresponding to tube diameters 
of $0.89nm$ and $1.49nm$ and resulting interlayer spacing 0.3nm. The spectrum at the top of figure 2 
shows several RB modes at higher Raman shifts ($336cm^{-1}$, $342cm^{-1}$ and $368cm^{-1}$). The RB 
modes in this range correspond to tubes of diameter $0.666nm$, $0.653nm$, $0.610nm$ which are particularly 
small and can be assigned to chiralities (8,1), (7,2) and (5,4). They can be associated to double walled 
tubes with RB modes of the external tube at $180cm^{-1}$ and the deduced interlayer spacing varies between 
$0.31-0.34nm$ for RB modes between $170-185cm^{-1}$. 

While the peaks at $336cm^{-1}$ and $368cm^{-1}$ fall within $3cm^{-1}$ of the predicted values for tubes with 
chirality (8,1) and (5,4), the peak at $343cm^{-1}$ falls between the values for tubes with chirality (8,1) 
and (7,2) (Table I) and could be associated with either of the two. In fact an increased bond length due to 
the higher curvature would shift the RB modes to lower energy but this should also be observed for the other 
tubes in the same diameter range. A reduced interlayer spacing would lead to a compression of the internal 
tube in a DWCNT and could explain the upshift of the RB mode. 

Recently, doubled walled CNT's have been synthesized through high temperature annealing ($1200^{o}C$) of 
single walled tubes filled with C$_{60}$ \cite{ijima}. The Raman spectra using the $514.5nm$ excitation of 
these tubes showed additional bands at $260cm^{-1}$, $315cm^{-1}$ and $377cm^{-1}$ with line widths of 
$>8cm^{-1}$. The diameter difference of inner and outer diameter (0.71nm) or interlayer distance (0.36nm) 
was found to be systematically larger than in graphite (0.335nm) in contrast to our findings here on 
catalytic CVD grown DWCNTs. 

A systematic up-shift of the RB modes of single walled CNT's has been associated to intertubular interaction and
distortion in bundles \cite{uma} and theoretical calculations indicate a up-shift of upto 10\% for tubes in 
bundles \cite{bern}. Furthermore a theoretical study has recently shown that gas absorption within the 
interstitial channels of a bundele of CNTs can increase the RB mode energy \cite{cole,ebbesen} by 2.7\%. This 
can certainly influence the position of the RB modes of the external tunbes but less likely the RB modes of the 
internal tubes. Taking into account the influence of the neighboring tubes on the RB mode of the external tube 
we find that the deduced interlayer spacing is larger ($0.32nm$) and brings the interlayer spacing closer 
to graphite. 

Table I shows that the differences between the experimental and theoretical values of the RB modes are 
different for semiconducting and metallic tubes. This difference can be explained by resonace effects,
sensitive to changes in the electronic structure of the tube. Inelastic light spectra of single shell 
carbon nanotubes \cite{tan} and of our DWCNT taken with an excitation source at $488.0nm$ have been found 
to be downshifted compared to spectra taken at $514.5nm$ and this can explain the differences observed for two 
types of DWCNT's. But the deduced different interlayer distance in the two types of double walled tubes could 
also indicate differences in the growth mechanism of catalytic CVD grown and annealed doubled walled tubes. 
Annealing causes a growth of an internal tube whereas in the CVD method presumably the external tube grows 
on the inner tube. This is consistent with the tendency that the layer grown later in the process forms 
before an interlayer distance corresponding to the value observed in graphite is reached.  

Among the observed chiralities of the internal tubes we have found neither a zigzag (n,0) nor a armchair 
(n,n) tube and in the diameter range of the internal tubes we find among the chiralities which have not 
been observed, three semiconducting and three metallic tubes. This fact indicates that the growth of 
the tubes with chiralities other than zigzag or armchair may be favored due to their lower symmetry.
Finally we note the particlularly narrow line shape of some of the RB modes of the internal tubes which in 
fact made it possible to assign the chirality of the tubes. The dispersion of the tubes on the bilayer 
substrate and selection of a single tube bundle using a microscope has the effect to reduce the broadening 
due to size dispersion. The narrow line shape shows the high coherence length and harmonicity for RB modes 
of internal tubes. The RB modes are localized on a very narrow tube and we expect that DWCNT's are as a 
result exceptionally good thermal conductors.



We have observed RB modes of DWCNTs by inelastic light scattering. The narrow diameter of the internal tubes 
grown by the catalytic CVD method, the higher sensitivity of the RB mode for smaller diameter and the 
reduced number of chiral vectors in this diameter range as well as the narrow line widths made it possible 
to assign the chirality of the internal tubes. We observe two metallic as well as several semiconducting tubes
and no zigzag (n,0) or armchair (0,m) tubes. Differences between experimental and theoretical energies of the 
RB modes are found to be larger for semiconducting tubes and attributed to mostly electronic resonance 
effects. 

Acknowledgement:

We would like to thank A. Zwick for helpful discussions and R. Sirvin for technical support.

Table\\
\\
\begin{table}
\caption{Experimental and predicted RB energies}
\begin{tabular}{c|c|c|c|c}
\hline
experiment &theory &difference &diameter &chirality\\
\hline
$cm^{-1}$&$cm^{-1}$&$cm^{-1}$&$nm$&$(n,m)$\\
\hline
297.0&296.2&0.8&0.756&(7,4) metallic\\
\hline
302.0&299.4&2.6&0.747&(9,1),(6,5) semiconducting\\
\hline
312.0&311.6&0.4&0.718&(8,2) metallic\\
\hline
336.0&334.3&1.7&0.669&(8,1) semiconducting\\
\hline
342.0&348.9&-6.9&0.641&(7,2) semiconducting\\
\hline
368.0&365.7&2.3&0.612&(5,4) semiconducting\\
\hline
\end{tabular}
\end{table}

\newpage

Figure Captions\\
\\
\begin{itemize}
\item[1.] Room temperature Raman spectra of RB modes on two location of the sample. The spectrum at the 
bottom is assigned to mostly single walled tubes and the spectrum at the top can be assigned to 
mostly double walled carbon nanotubes.

\item[2.] Room temperature Raman spectra of RB modes along a tube bundle containing 
double walled carbon nanotubes with different internal and external diameters.

\end{itemize}


\begin{thebibliography}{99}
\bibitem{stat} Statistical analysis on 67 isolated nanotubes of the same sample showed that 38$\%$ of the tubes 
are single walled, 58$\%$ double walled and the remaining having three or four walls. In the smallest diameter 
range (0.8-1.0nm) we find only double walled tubes.
\end{thebibliography}
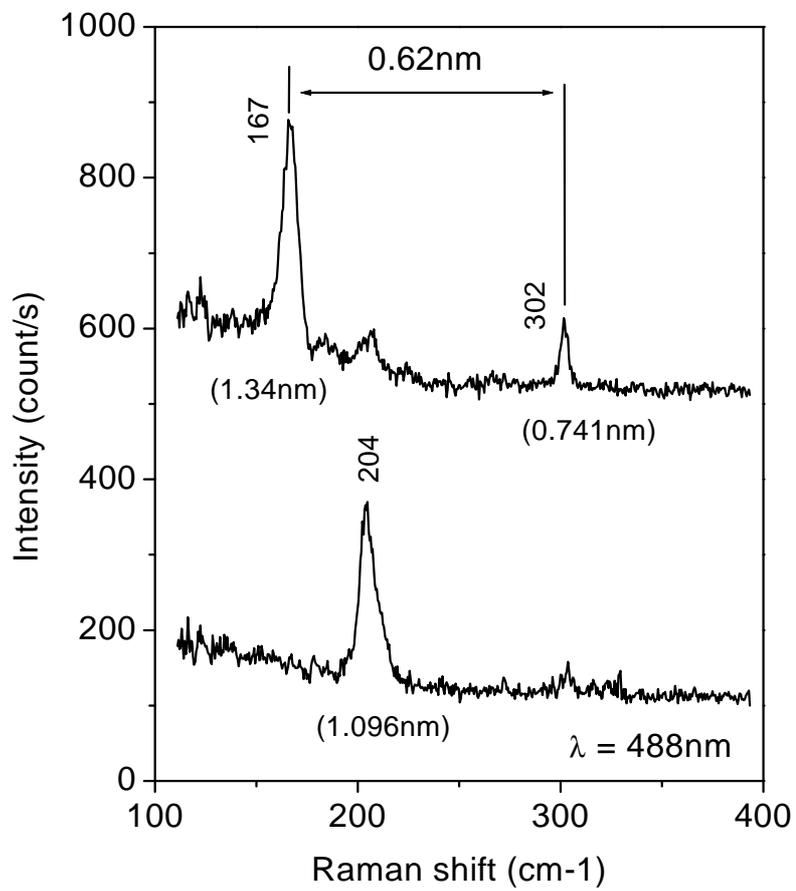
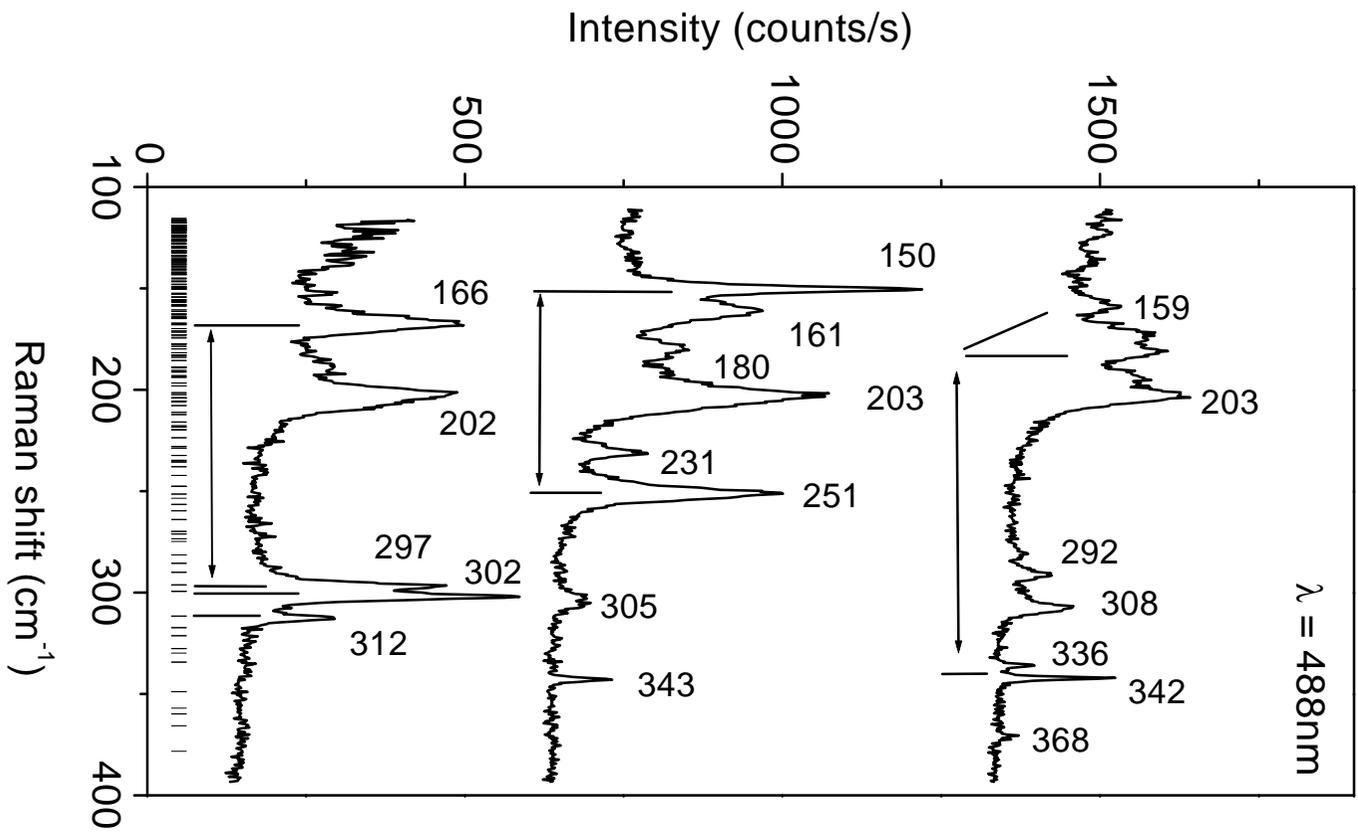
\end{document}